\documentclass[aps,prl,amsmath,amssymb,amsfonts,showpacs,superscriptaddress,floatfix,twocolumn]{revtex4}

\usepackage{graphicx}
\usepackage{epsfig}
\usepackage{braket}

\def\be{\begin{equation}}
\def\ee{\end{equation}}
\def\f#1#2{\frac{#1}{#2}}
\def\bea{\begin{eqnarray}}
\def\eea{\end{eqnarray}}

\begin{document}

\title{Entanglement rates and area laws}

\author{Karel Van Acoleyen}
\affiliation{Department of Physics and Astronomy,
Ghent
 University,
  Krijgslaan 281, S9, 9000 Gent, Belgium}

\author{Micha\"{e}l Mari\"{e}n}
\affiliation{Department of Physics and Astronomy,
Ghent
 University,
  Krijgslaan 281, S9, 9000 Gent, Belgium}

  \author{Frank Verstraete}
  \affiliation{Department of Physics and Astronomy,
Ghent
 University,
  Krijgslaan 281, S9, 9000 Gent, Belgium}
  \affiliation{Vienna Center for Quantum Science and Technology, Faculty of Physics, University of Vienna, Boltzmanngasse 5, 1090 Vienna, Austria}

\begin{abstract}
We prove an upper bound on the maximal rate at which a Hamiltonian interaction can generate entanglement in a bipartite system. The scaling of this bound as a function of the subsystem dimension on which the Hamiltonian acts nontrivially is optimal and is exponentially improved over previously known bounds. As an application, we show that a gapped quantum many-body spin system on an arbitrary lattice satisfies an area law for the entanglement entropy if and only if any other state with which it is adiabatically connected (i.e. any state in the same phase) also satisfies an area law.
\end{abstract}

\maketitle

\noindent \emph{Introduction} Entanglement is one of the defining trademarks of quantum mechanics, appearing ubiquitously, both at the theoretical and experimental level.
For many of the applications, notably in quantum optics, nuclear magnetic resonance and condensed matter physics, it is the optimal \emph{creation} of entanglement that is important, and much experimental effort has been devoted to this process. A fundamental question then is, given some Hamiltonian interaction between two subsystems $A$ and $B$, what is the maximal rate at which the Hamiltonian evolution can generate entanglement \cite{Dur,Bravyi,Bennett}? It is this \emph{dynamical} question that we study in this paper, and we will provide a tight upper bound for the maximal entanglement rate in the most general setting. But as we will show in the second part of the paper, this issue also has important consequences for the \emph{static} entanglement properties of quantum many-body systems.

Over the past decade it has been realized that looking at entanglement properties provides a new window on these systems, allowing an improved classification of different quantum phases of matter \cite{Chen,Schuch,Sachdev}. In particular it was found that the entanglement distribution for gapped systems always seems to have the same characteristic behavior \cite{Eisert}: for two arbitrary connected subsystems, the entanglement entropy scales like the boundary area (with small logarithmic corrections for critical systems), rather than the volume as one would expect for random states. This so called area law has important consequences for the efficient representability of quantum many-body states. And as such it has served as a guiding principle for the formulation of the different successful numerical tensor network methods that have emerged in the last years \cite{Cirac}. However, so far the area law has only been proven for gapped one-dimensional systems \cite{Hastings, Arad}. Our results provide a step in the direction of a general proof for higher dimensions. Using the formalism of quasi-adiabatic continuation \cite{HastingsWen}, our bound on the entanglement rate allows us to show that within a gapped quantum phase,  a subsystem's entropy is constant up to a term that scales like its boundary area. This then implies that an area law for one particular system automatically carries through to all other systems in the same quantum phase.   A similar statement was made in the concluding remarks of \cite{Bravyi II}, but it turns out that the argument presented there was not strong enough as it only applied to adiabatic evolution and not to quasi-adiabatic as is needed for proving area laws.  Also, this stability of the area law under adiabatic evolution was claimed by Michalakis in \cite{Michalakis}, but he needed extra assumptions on the decrease of the Schmidt coefficients and his results also contain an extra prefactor which scales superlogarithmically with the area \cite{footnote II}.\\

Let us now first return to the problem of bounding the maximal dynamical entanglement rate for bipartite systems. We consider a pure state $\ket{\Psi}$ on a system $aABb$, for which the time evolution is governed by the unitary operator $U(t)=e^{i(H_{aA}+H_{bB}+H_{AB})t}$. The entanglement entropy between subsystems $aA$ and $Bb$ can be quantified by the von Neumann entropy, $S_{aA}(t)=-Tr\,\rho_{aA}(t)\log \rho_{aA}(t)$, with $\rho_{aA}(t)=Tr_{\!Bb}U(t)\ket{\Psi}\bra{\Psi}U(t)^\dag$. Accordingly, the entanglement rate is then defined as: \be \Gamma=\f{dS_{aA}(t)}{dt}|_{t=0}\,. \label{rate}\ee
Notice that we are explicitly considering ancillas $a$ and $b$ that do not directly interact with the other subsystem. But although $H_{aA}$ and $H_{bB}$ do not contribute to the entanglement rate (\ref{rate}), the ancillas can still influence this rate indirectly, through their entanglement with the rest of the system. It is precisely the influence of these ancillas that we address in this paper. Notice also that we are looking for a bound on the entanglement rate at some particular (arbitrary) reference time, as opposed to a bound on the average rate over some period.

The problem of maximizing $\Gamma$ was first considered in \cite{Dur}, for the case where $A$ and $B$ are qubits ($d_A=d_B=2$). It was shown that, in the absence of ancillas, $\Gamma_{max}\equiv\max_{\Psi}\Gamma \leq \beta ||H||$, with $||H||$ the operator norm of the interacting Hamiltonian $H_{AB}$ and $\beta\approx 1.9123$. Furthermore the authors observed that ancillas can generically increase the maximal entanglement rate. The general case with no ancillas was solved by Bravyi in \cite{Bravyi}, who obtained $\Gamma_{max}\leq c(d)||H||\log d$. Here $d=\min(d_A,d_B)$, the smallest dimension of the interacting subsystems. The constant $c(d)$ decreases with $d$, with $c(2)=\beta$ and $c(d)\rightarrow 1$ for large $d$. (See also \cite{Hutter} for an alternative proof that $c\leq 4$ in the absence of ancillas.) Remark that these bounds are optimal, one can always find a particular $H_{AB}$ for which they are saturated \cite{Dur,Bravyi}.

In \cite{Bravyi} the small incremental entangling (SIE) conjecture, attributed to  Kitaev, was put forward for the case of ancilla-assisted entanglement rates:
 \be \Gamma_{max}\leq c ||H||\log d\quad\quad d=\min(d_A,d_B)\,,\label{Kitaev}\ee
where $c$ is an order one constant independent of $d$. Bennett et al had already shown in \cite{Bennett} that the ancilla-assisted entanglement rate is indeed bounded by a quantity that is independent of the ancilla dimensions, obtaining $\Gamma_{max}\leq c ||H|| d^4$. The results in \cite{Bravyi} imply a stronger bound $\Gamma_{max}\leq 2 ||H|| d^2$ and very recently this bound was further improved by Lieb and Vershynina \cite{Lieb}, obtaining $\Gamma_{max} \leq (4/\ln2) ||H|| d$. But for large systems, this last bound is still exponentially weaker than the conjectured bound (\ref{Kitaev}). To prove our result on the adiabatic evolution of the entanglement entropy within a many-body quantum phase, one needs the stronger bound (\ref{Kitaev}), which will be proven now.

\noindent \emph{Proof of the SIE conjecture}: W.lo.g. we will work with a normalized interaction Hamiltonian $H_{AB}$ with norm $||H||=1$. To prove SIE we will start by following the strategy that was set out in \cite{Bravyi}. If we consider the case where $d_A\geq d_B$ the bound (\ref{Kitaev}) that we want to prove reads: $\Gamma \leq c\log d_B$. As the right-hand side of this bound is independent of $a$ and $A$, we can simply consider the case $d_a=1$, indeed, we can always extend $A$ to $A\otimes a$. The entanglement rate then reads: \be \Gamma=-i\,Tr\left(H_{AB}[\rho_{AB},\log \rho_A\otimes I_B]\right)\,, \ee which we can formally recast as: \be \Gamma=\f{1}{p}\Lambda(p)\equiv\f{1}{p}\left(-i\,Tr\left(H[X,\log Y]\right)\right)\,, \label{deflambda}\ee identifying $H=H_{AB}$, $X=\rho_{AB}/d_B^2$, $Y=\rho_A \otimes I_B/d_B$ and $p=1/d_B^2\leq1/2$, and using that the commutator remains unchanged upon inclusion of the constant factor $1/d_B$ in the
 log. The idea now is to bound $\Lambda(p)$ \cite{footnote} for general $H,X,Y$, under the conditions: \be ||H||=1, \quad TrX=p,\quad TrY=1\quad  0\leq X\leq Y\,.\label{conditions}\ee These conditions indeed follow from the identification above, in particular the last inequality follows from $\rho_{AB}\leq d_B \rho_A\otimes I_B$, as one can easily show (see \cite{Bravyi}). It is clear then, that if we prove that for any matrix dimension, $\Lambda(p)\leq -c' p\log p$, we will have proven the SIE conjecture (\ref{Kitaev}) with $c=2c'$. We will obtain $c=18$.

Let us first do the optimization over $H$, which can be cast into a variational problem over all possible projectors $0\leq P\leq I$:
\be \max_{||H||=1}|\Lambda(p)|=2\max_P\,|\,Tr\left(P[X,\log Y]\right)|\,. \label{maxH}\ee

Working in the eigenbasis of $Y$, the quantity $|\Lambda(p)|$ we want to bound then reads: \be 2|\sum_{i<j}\log\f{y_i}{y_j}(X_{ij}P_{ji}-X_{ji}P_{ij})|\,,\label{SIM}\ee
with $P$ some projector and the nonzero eigenvalues $y_i$ of $Y$ in decreasing order: $y_1\geq y_2\geq \ldots \geq y_N>0$\,.

It is now useful to group these $N$ eigenvalues in a finite number of successive intervals:
\bea 1> y_{i_1}\geq p &\quad \quad& 1\leq i_1\leq  n_1\nonumber\\
 p> y_{i_2}\geq p^2 &\quad \quad& n_1< i_2\leq  n_2\nonumber\\
 \ldots \nonumber\\
 p^{k-1}>y_{i_k}\geq p^k &\quad\quad& n_{k-1}< i_k\leq  n_k \label{intervals}\\
 \ldots\nonumber\\
 p^{k^*-1}>y_{i_{k^*}}\geq p^{k^*}&\quad\quad& n_{k^*-1}<i_{k^*}\leq N\nonumber\eea
 Of course some intervals can be empty, in this case we have $n_{k-1}=n_k$. We can now rearrange the sum in (\ref{SIM}) as follows (with $i_k, j_k \!\in \,]n_{k-1},n_k]$):  \bea \sum_{i<j}&=&(\sum_{i_1<j_1}+\sum_{i_1,i_2}+\sum_{i_2<j_2})+(\sum_{i_2<j_2}+\sum_{i_2,i_3}+\sum_{i_3<j_3})\nonumber\\
 &&+\ldots +(\!\!\!\!\!\!\sum_{i_{k^*-1}<j_{k^*-1}}+\sum_{i_{k^*-1},i_k^*}+\sum_{i_k^*<j_k^*}) \nonumber\\
 &&-(\sum_{i_2<j_2})-(\sum_{i_3<j_3})-\ldots -(\sum_{i_{k^*-1}<j_{k^*-1}})\nonumber\\
 &&+(\sum_{i_1,i_{k>2}}+\sum_{i_2,i_{k>3}}+\ldots+\sum_{i_{k^*\!-2},i_{k^*}})\,.\label{rearrange}\,\eea
Our bound then follows from separately bounding the absolute values of all individual bracketed terms in this sum.  This of course leads to a bound on the absolute value of the full sum in (\ref{SIM}). 

The logic behind the particular rearrangement (\ref{rearrange}) is twofold. First of all, the sum on the last line now only runs over pairs $(i,j)$ that are separated by at least one interval in (\ref{intervals}). For any pair in this restricted sum $\tilde{\sum}$ we therefore have: $y_j<py_i$, which allows for a useful bound in the following way. We write $X$ as: $X=Y^{1/2}ZY^{1/2}$ with $0\leq Z\leq I$, as follows from (\ref{conditions}). The contribution to the bound of (\ref{SIM}) then reads:
\be 2|\tilde{\sum_{i<j}}\log\f{y_i}{y_j}y_i^{1/2}y_j^{1/2}(Z_{ij}P_{ji}-Z_{ji}P_{ij})|\,,\ee
which from the Cauchy-Schwarz relation can be bounded by: \bea &&4(\tilde{\sum_{i<j}}\log\f{y_i}{y_j}y_i^{1/2}y_j^{1/2}Z_{ij}Z_{ji})^{\f{1}{2}} (\tilde{\sum_{i<j}}\log\f{y_i}{y_j}y_i^{1/2}y_j^{1/2}P_{ij}P_{ji})^{\f{1}{2}}\nonumber\\
&&\leq4p^{1/2}\log(1/p)(\tilde{\sum_{i<j}}y_iZ_{ij}Z_{ji})^{\f{1}{2}}\, (\tilde{\sum_{i<j}}y_iP_{ij}P_{ji})^{\f{1}{2}}\nonumber\\
&&\leq4p^{1/2}\log(1/p)(\sum_{i,j=1}^{N}y_iZ_{ij}Z_{ji})^{\f{1}{2}} \,(\sum_{i,j=1}^Ny_iP_{ij}P_{ji})^{\f{1}{2}}\nonumber\\
&&\leq4p^{1/2}\log(1/p)(\sum_{i=1}^Ny_iZ_{ii})^{\f{1}{2}} \,(\sum_{i=1}^Ny_iP_{ii})^{\f{1}{2}}\nonumber\\
&&\leq4p\log(1/p)\,.\eea
On the second line we use $x^{1/2}\log(1/x)\leq p^{1/2}\log(1/p)$, if $x\leq p \leq 1/e^2$, with $x=y_j/y_i$. On the third line we simply add positive terms in the summation, while on the fourth line we use $Z,P\leq I$. Finally on the fifth line we use $p=Tr(X)=Tr(YZ)$ and $Tr Y=1$.

The other terms on the first three lines in (\ref{rearrange}) are grouped such that any bracketed term can be rewritten again as a matrix expression, now on a subspace corresponding to two subsequent intervals in (\ref{intervals}) (line one and two) or to a single interval (line three). This then also allows for a useful bound. For the first term on line one for instance we have:
\bea &&2|\sum_{i=1}^{n_2}\sum_{j=i+1}^{n_2}\log\f{y_i}{y_j}(X_{ij}P_{ji}-X_{ji}P_{ij})|\nonumber\\
&&=2|Tr \tilde{P}[\tilde{X},\log\tilde{Y}]|\,\nonumber\\
&&\leq ||[\tilde{X},\log\tilde{Y}]||_1\nonumber\\
&&=||[\tilde{X},\log\left(\tilde{Y}/\tilde{y}_{min}\right)]||_1\label{part}\\
&&\leq ||\log\left(\tilde{Y}/\tilde{y}_{min}\right)||\,||\tilde{X}||_1\nonumber\\
&&=\log\f{\tilde{y}_{max}}{\tilde{y}_{min}}\, Tr \tilde{X}\nonumber\\
&&\leq2\left(p_1+p_2\right)\log(1/p)\,.\nonumber\eea
Here, on the second line $\tilde{Y}$ denotes the part of $Y$ acting on the subspace spanned by the eigenvectors $1$ to $n_2$, while $\tilde{P}\leq I$ and $\tilde{X}$ are two semi-positive definite operators on this subspace. On the third line we use that the maximum will be reached for $\tilde{P}$ a projector, yielding the trace-norm of the commutator. On line four we add a constant factor $1/\tilde{y}_{min}$ in the log, leaving the commutator unchanged.   With $\tilde{y}_{min}$ defined as the minimal eigenvalue of $\tilde{Y}$, the right-argument of the commutator is now also semi-positive. This allows us to use the commutator inequality by Kittaneh \cite{Kittaneh} on line five. Finally on the last line we use that $p^2<y_i/y_j<1/p^2$ for $y_i, y_j$ elements of two neighboring intervals (\ref{intervals}), and we also write $Tr\tilde{X}=\sum_{i=1}^{n_2} X_{ii}\equiv p_1+p_2$. Defining the full decomposition of $p
 =Tr X$ as: \be p=\sum_{k}\sum_{i_k} X_{i_ki_k}\equiv \sum_k p_k\,,\ee we then find the contribution of the full first and second line in (\ref{rearrange}) to the bound of (\ref{SIM}):  \be 2(p_1
  + p_{k^
 *}+2\sum_{k=2}^{k^*-1} p_k)\log\f{1}{p}\leq 4 \sum_{k=1}^{k^*} p_k\log\f{1}{p}=4p\log\f{1}{p}\,. \ee  In a similar fashion we can bound the contribution of the third line by $p\log(1/p)$. Taking the contributions of the four lines in (\ref{rearrange}) together, we finally find the bound (for $p\leq1/e^2$): \be \Lambda(p) \leq 9p\log(1/ p)\,,\label{bound}\,\ee which concludes our proof of the SIE conjecture (\ref{Kitaev}), with $c=18$ (and for $d=\sqrt{1/p}\geq 3$) $\Box$.

This is an improvement of the bound in \cite{Lieb} only for $d>10$. But we now indeed recover the optimal $\log d$ scaling of the SIE conjecture (\ref{Kitaev}). Our pre-factor $c=18$ itself is probably not optimal. Bravyi put forward the SIM conjecture, $\Lambda(p)\leq c"(-p\log p-(1-p)\log(1-p))$ \cite{Bravyi}, and found $c"=1$ on numerical examples.  We also seem to find this numerically for matrix dimensions up to 1000. For large $d$ this would lead to $\Gamma_{max} \lesssim 2 ||H|| \log d$.

\noindent\emph{Area law for quasi-adiabatic continuation:} Two gapped systems are defined to be in the same phase if and only if they can be connected by a smooth path of gapped local Hamiltonians \cite{ChenII}.
More explicitly, if we parameterize this path with $s\in[0,1]$, all ground states $\Ket{\Psi(s)}$ of the continuous family of gapped local Hamiltonians $H(s)$ then belong to the same phase as $\Ket{\Psi(0)}$. We can now employ the bound (\ref{Kitaev}) to  put a bound on the variation of some subsystem's entanglement entropy, along the path traced out by $s$. The formalism of quasi-adiabatic continuation, first introduced in \cite{HastingsWen}, permits us to write a Schr\"{o}dinger equation for the \emph{exact} evolution along this path: \be \f{d}{ds}\ket{\Psi(s)}=iK(s)\ket{\Psi(s)}\,. \ee The power of this formalism lies in the fact that for local gapped Hamiltonians $H(s)$, one can show that the effective Hamiltonian $K(s)$ will in fact be \emph{quasi-local} \cite{Osborne,Bachmann}. To be specific (and from now on we largely follow the notation of \cite{Osborne}), let us write the Hamiltonian as a sum of local near(est) neighbor interactions, $H(s)=\sum_{\bf{i} \in \lambda}
  h_{\bf{
 i}}(s)$,  on a $D$-dimensional lattice $\lambda$. For simplicity we will consider a translation invariant system, with $||h_{\bf{i}}(s)||= ||h_0(s)||$. The corresponding effective Hamiltonian then reads \cite{Osborne}: $K(s)=\sum_{\bf{i} \in \lambda} k_{\bf{i}}(s)$. Here, each term $k_{\bf{i}}$ now has support on the full lattice, but the interaction strength decays for large distances. That is, we can write this term as: \be k_{\bf{i}}(s)= \sum_{r=0} k_{\bf{i}}(s,r)\,, \ee where $k_{\bf{i}}(s,r)$ has support on a ball with radius $r$, centered at the point $\bf{i}$, and its magnitude $||k_{\bf{i}}(s,r)||=||k_0(s,r)||$ decays sub-exponentially with $r$, either polynomial or superpolynomial, depending on the specific choice for the filter function in $K(s)$.

\begin{figure}[t]
\centering
\includegraphics[width= 7.5cm]{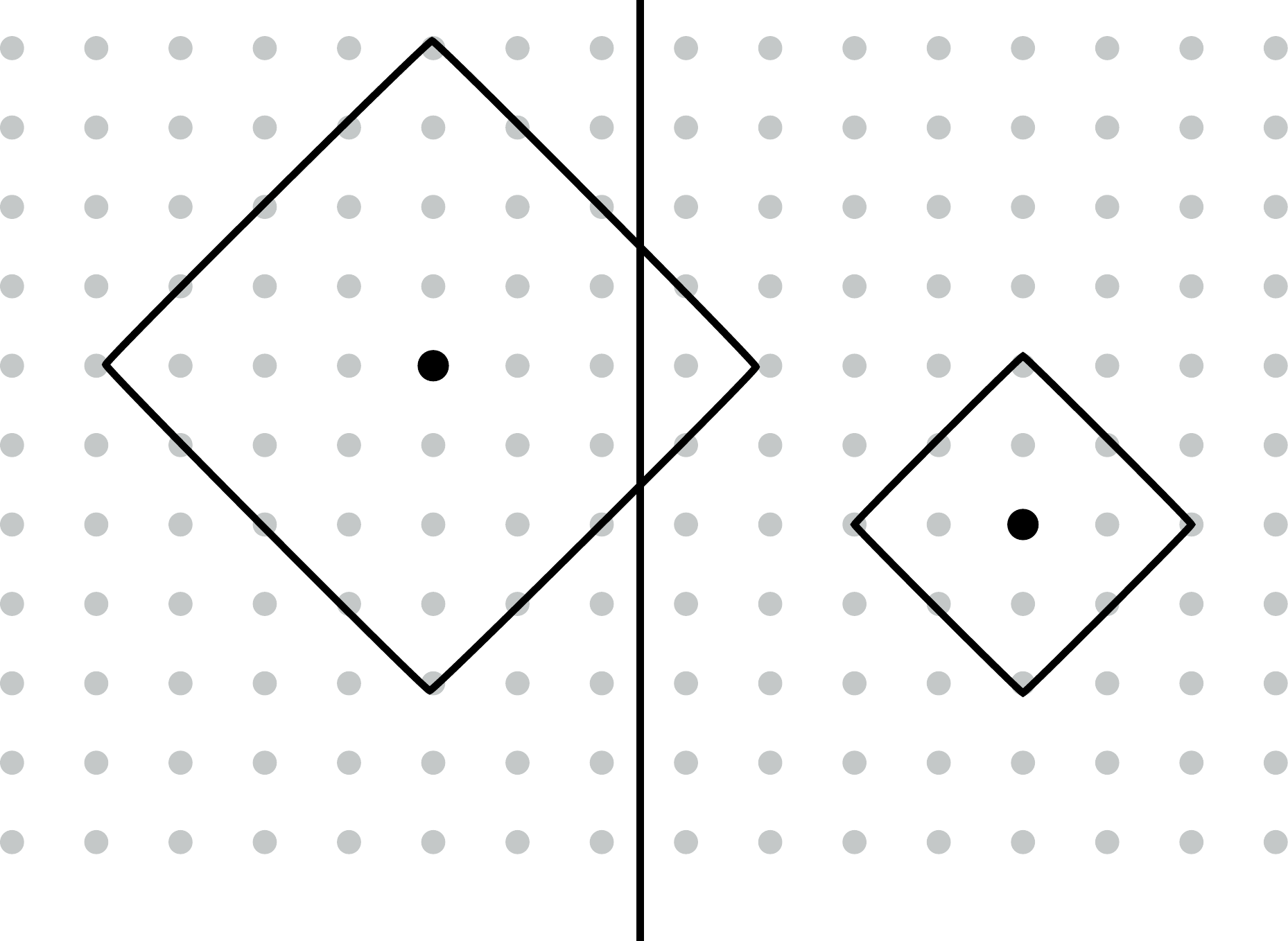} \label{figuur}
\caption{The balls of support (for the Manhattan metric) centered at two points $\bf{i}$ and $\bf{j}$, for two terms $k_{\bf i}(s,4)$ and $k_{\bf j}(s,2)$ in the Hamiltonian $K(s)$. Only the term $k_{\bf{i}}(s,4)$ can generate entanglement across the cut.}
\end{figure}
If we now take a bipartition of the full system into $L$ and $R$, and for the sake of clarity we consider a straight cut (see Fig.1),  we can bound the entanglement rate along $s$ as follows (working in lattice units $a=1$, and dropping order one pre-factors): \bea \f{dS_L(s)}{ds}&=&i\,Tr(K(s)[\,\ket{\Psi(s)}\Bra{\Psi(s)},\log \rho_L\otimes I_R])\nonumber\\
&=&i\,Tr (\sum_{{\bf i},r} k_{\bf{i}}(s,r)[\,\ket{\Psi(s)}\Bra{\Psi(s)},\log \rho_L\otimes I_R])\nonumber \\
&\lesssim&  \sum_{({\bf i}_P,i_O)}\sum_{r\geq i_O} \log d_{l}\, r^D ||k_{{\bf i}_P,i_O}(s,r)||\label{adrate}\\
&\lesssim&\log d_l\, A\, ( \sum_r r^{D+1} ||k_0(s,r)||)\nonumber\,.  \eea
On line three we decompose ${\bf i}=({\bf i}_P,i_O)$ in the directions parallel and the direction orthogonal to the boundary and we apply the bound (\ref{Kitaev}) for each term $k_{\bf{i}}(s,r)$ separately. We overestimate $d$ in (\ref{Kitaev}) by equating it to the full dimension of the subsystem where $k_{\bf{i}}(s,r)$ is working on: $\log d \approx \log d_l r^D$, with $d_l$ the local dimension (and dropping the geometric volume pre-factor). The restriction on the sum over $r$ follows from the fact that only the terms that have overlap in both subsystems $L$ and $R$ can generate entanglement. Finally on the last line, the sum over the parallel indices gives us the boundary area $A$ and we have summed the $i_O$ index. It is clear then that we find an area law for the entanglement rate along quasi-adiabatic continuation, as long as  $||k_0(s,r)||$ decays faster than $1/r^{D+2}$. Specifically, with a little effort one can read off the following bounds from \cite{Osborne} (up to order one pre-factors) for
 the case of a polynomially decaying filter function: \bea ||k_0(r,s)||&\lesssim&\f{||h'||}{\gamma}\,,\nonumber\\
||k_0(r,s)||&\lesssim& r^{D-1}\f{||h||||h'||\gamma}{\kappa^3}e^{-r v/2}+\f{||h'||}{\gamma}\left(\f{\xi}{r}\right)^{n}\,,\label{kbounds}\eea
with $||h||=||h_{0}(s)||$ and $||h'||=||dh_{0}(s)/ds||$. $\gamma(s)$ is the mass gap, $\kappa$ and $v$ are two constants that appear in the Lieb-Robinson bound \cite{LiebRobinson}, from which the Lieb-Robinson velocity reads $v_{LR}=\kappa/v$ , and $\xi=v_{LR}/\gamma$ is the correlation length that follows for gapped systems \cite{HastingsII}. In the case of nearest-neighbor interactions, one can choose the Lieb-Robinson constants as: $\kappa\approx ||h||$ and $v=1$ . Furthermore, as we mentioned before, the power of decay $n$, depends on the specific form of $K(s)$ and can be chosen freely. For $n>D+2$, we can then bound the sum in the entanglement rate (\ref{adrate}), by using the first bound for $r\leq\xi$ and the second bound for $r>\xi$. For the physically interesting case $\gamma\leq ||h||$, or equivalently $\xi\geq 1$, the second term of this second bound will dominate, and we find:
\be \sum_r r^{D+1} ||k_0(s,r)||\lesssim \f{||h'||}{\gamma}\xi^{D+2}\,,\ee
resulting in the area law: \be \f{dS_L(s)}{ds}\lesssim A\f{||h'(s)||}{\gamma(s)}\xi(s)^{D+2} \log d_l , \label{areaad}\ee
for the variation of the subsystem entanglement entropy along an adiabatic path. Upon integration of (\ref{areaad}) we can then conclude
$\Delta S_L=S_L(s)-S_L(0)\leq A \tilde{c}(s) \log d_l $, for the subsystem entropy difference for two states belonging to the same gapped quantum phase. As $\tilde{c}$ is independent of the system size or boundary area $A$, we have indeed shown for the first time that an entropy area law for one gapped system implies an area law for all systems within the same quantum phase.\\

 Notice that the formalism of quasi-adiabatic evolution also works in the case of topological quantum phases \cite{Bravyi III}; from (\ref{areaad}) we can then indeed conclude that an area law for the groundstates is protected under adiabatic evolution within a gapped topological quantum phase.\\

In conclusion, we have proven an upper bound on the entanglement that can be generated by any Hamiltonian which acts on a subsystem of a large bipartite system, originally conjectured by Bravyi and Kitaev. The corresponding bound is optimal to within a constant, and its scaling w.r.t. the dimension of the subsystem on which it acts nontrivially is logarithmic. Our motivation for proving this bound was to understand the scaling of the entanglement entropy in ground states of quantum many-body systems within the same phase. By combining this small incremental entangling theorem with Lieb-Robinson techniques, we were able to prove that a ground state of a quantum spin system obeys an area law if and only if all other ground states in the same phase obey an area law, and this result is valid in any dimension and on any lattice. We hope that this might present a first step in proving the existence of an area law in any gapped system in higher dimensions.

During completion of our manuscript, Audenaert constructed an alternative proof of the SIM conjecture (implying the SIE conjecture) \cite{Audenaert}.

\acknowledgments We would like to acknowledge J.I. Cirac for convincing us that Cauchy-Schwarz was the way to go, K. Audenaert for making us aware of the Kittaneh inequality and K. Audenaert, I. Cirac, B. De Vylder, E. Lieb, S. Michalakis and T. Osborne for extremely inspiring and useful discussions. This work is supported by an Odysseus grant from the FWO,  the FWF grants FoQuS and Vicom, and the ERC grant QUERG.


\begin{thebibliography}{99}


\bibitem{Dur} D{\"u}r, W., Vidal, G.,
Cirac, J.~I., Linden, N.,
\& Popescu, S.\ 2001, Physical Review Letters, 87, 137901


\bibitem{Bravyi} Bravyi, S.\ 2007, \pra, 76,
052319

\bibitem{Bennett} Bennett, C.~H., Harrow,
A.~W., Leung, D.~W., \& Smolin, J.~A.\ 2002, IEEE Trans. Inf. Theory, Vol. {\bf 49}, No. {\bf 8}, p. 1895 (2003), [arXiv:quant-ph/0205057].

\bibitem{Chen} Chen, X., Gu, Z.-C.,
\& Wen, X.-G.\ 2011, \prb, 84, 235128

\bibitem{Schuch} Schuch, N.,
P{\'e}rez-Garc{\'{\i}}a, D., \& Cirac, I.\ 2011, \prb, 84, 165139


\bibitem{Sachdev} Sachdev, S.\ 2012, 25th Solvay Conference on Physics,
[arXiv:1203.4565]

\bibitem{Eisert} Eisert, J., Cramer, M.,
\& Plenio, M.~B.\ 2010, Reviews of Modern Physics, 82, 277

\bibitem{Cirac} Cirac, J.~I., \& Verstraete, F.\ 2009, Journal of Physics A Mathematical General, 42, 4004

\bibitem{Hastings}Hastings, M.~B.\ 2007,
Journal of Statistical Mechanics: Theory and Experiment, 8, 24

\bibitem{Arad} Arad, I., Kitaev, A.,
Landau, Z., \& Vazirani, U.\ 2013, arXiv:1301.1162

\bibitem{HastingsWen} Hastings, M.~B., \& Wen, X.-G.\ 2005, \prb, 72, 045141

\bibitem{Bravyi II} Bravyi, S., Hastings,
M.~B., \& Verstraete, F.\ 2006, Physical Review Letters, 97, 050401

\bibitem{Michalakis} Michalakis, S.\ 2012,
arXiv:1206.6900

\bibitem{footnote II}  $R_0$ in eq. (23) of \cite{Michalakis} depends (superlogarithmically) on $|\delta A|$ through eq. (20).

\bibitem{Hutter}
 Hutter, A., \& Wehner, S.\ 2012, Physical Review Letters, 108, 070501


\bibitem{Lieb} Lieb, E.~H., \& Vershynina, A.\ 2013, arXiv:1302.3865

\bibitem{footnote} $\Lambda(p)$ can be formally identified as a small incremental mixing rate (SIM), describing the entanglement rate for a probabilistic ensemble of two mixed states \cite{Bravyi,Lieb,Audenaert}.

\bibitem{Kittaneh} F. Kittaneh,  J. Funct. Anal. 250 (2007) 132-143.

\bibitem{ChenII} Chen, X., Gu, Z.-C.,
\& Wen, X.-G.\ 2010, \prb, 82, 155138

\bibitem{Osborne} Osborne, T.~J.\ 2007, \pra,
75, 032321

\bibitem{Bachmann} Bachmann, S.,
Michalakis, S., Nachtergaele, B.,
\& Sims, R.\ 2012, Communications in Mathematical Physics, 309, 835

\bibitem{LiebRobinson}Lieb, E.~H., \& Robinson, D.~W.\ 1972, Communications in Mathematical Physics, 28, 251


\bibitem{HastingsII} Hastings, M.~B.\ 2004,
Physical Review Letters, 93, 140402

\bibitem{Bravyi III} Bravyi, S., Hastings,
M.~B., \& Michalakis, S.\ 2010, Journal of Mathematical Physics, 51, 093512

\bibitem{Audenaert} Audenaert, K.~M.~R.\ 2013,
arXiv:1304.5935









\end{thebibliography}
\end{document}